\documentclass[%
 reprint,
 amsmath,amssymb,
 aps,
floatfix,
]{revtex4-2}

\usepackage{graphicx}
\usepackage{dcolumn}
\usepackage{bm}
\usepackage{aas_macros} 
\usepackage{color} 

\begin{document}

\preprint{APS/123-QED}

\title{Measuring the cosmological density field twice:\\ A novel test of dark energy using CMB quadrupole}

\author{Kiyotomo Ichiki$^{1,2}$}
 \email{ichiki@a.phys.nagoya-u.ac.jp}
\author{Kento Sumiya$^1$}%
\affiliation{%
$^1$Graduate School of Science, Division of Particle and
Astrophysical Science, Nagoya University, Chikusa-Ku, Nagoya, 464-8602, Japan\\
$^2$Kobayashi-Maskawa Institute for the Origin of Particles and the
Universe, Nagoya University, Chikusa-ku, Nagoya, 464-8602, Japan
}%
\author{Guo-Chin Liu}
\affiliation{
Department of Physics, Tamkang University, Tamsui, New Taipei City 25137, Taiwan
}%


\date{\today}

\begin{abstract}
The scattering of cosmic microwave background (CMB) radiation in galaxy clusters induces polarization signals according to the quadrupole anisotropy in the photon distribution at the cluster location. This `remote quadrupole' derived from the measurements of the induced polarization provides an opportunity for reconstructing primordial fluctuations on large scales. We discuss that comparing the local CMB quadrupoles predicted by these reconstructed primordial fluctuations and the direct measurements done by CMB satellites may enable us to test the dark energy beyond cosmic variance limits.
\end{abstract}

\maketitle


\section{\label{sec:intro}Introduction}
After the firm discovery of the accelerating expansion of the universe from observations of distant--type Ia SNs \cite{1998AJ....116.1009R,1999ApJ...517..565P}, dark energy, which is responsible for the cosmic acceleration, has been one of the biggest mysteries in cosmology.  In the last 20 years, detailed observations of cosmic microwave background (CMB) anisotropies have provided us with much information about the universe, and the cosmological parameters in the standard cold dark matter model with a cosmological constant $\Lambda$ ($\Lambda$CDM) have been precisely determined \cite{2013ApJS..208...19H,2020A&A...641A...6P}.  The CMB is sensitive to dark energy through the integrated Sachs--Wolfe (ISW) effect, where the decay of the gravitational potentials due to the accelerating expansion of the universe generates energy fluctuations in the CMB photons passing through these potentials. However, the CMB constraint on dark energy-related parameters is weak because the ISW effect appears only on large scales in the CMB temperature anisotropy spectrum and suffers from sizable cosmic variance errors.

An interesting idea of reducing the cosmic variance errors associated
with large-scale fluctuations was proposed by Kamionkowski and Loeb
\cite{1997PhRvD..56.4511K}. They argued that the polarization of the CMB
photons scattered off the free electrons in a cluster of galaxies could
be used to reduce the cosmic variance because the polarization is
sensitive to the quadrupole anisotropy of the CMB's last scattering
surface viewed by the cluster \cite{2000PhRvD..62l3004S}. However, as
Portsmouth pointed out \cite{2004PhRvD..70f3504P}, the quadrupoles
viewed by distant clusters are largely correlated with the local
quadrupole viewed by us. Therefore, the cosmic variance associated with
the local quadrupole is not reducible by the Kamionkowski and Loeb
method.  However, cluster polarization measurements can still provide
information on large-scale fluctuations \cite{2006PhRvD..73l3517B,2007PhRvD..75j1302A}, which can be useful for
studying the ISW effects \cite{2003PhRvD..67f3505C,2004PhRvD..69b7301C,2005PhRvL..95j1302S},
the power asymmetry of CMB polarization and density field \cite{2018JCAP...04..034D}, and the reionization optical depth \cite{2018PhRvD..97j3505M}.

In our previous paper \cite{2016MNRAS.460L.104L}, we showed that cluster CMB polarization measurements could be used to estimate the initial density fluctuations on large scales and reconstruct the local quadrupole of the CMB from our viewpoint.  We can reconstruct our quadrupole with a few hundred clusters at $0<z<1$ if we know the quadrupole transfer function, that is, if we know the correct cosmological model.  In other words, if we assume a wrong cosmological model to reconstruct the CMB quadrupole from distant clusters, the reconstructed CMB quadrupole will be different from that observed by the CMB satellites (e.g., WMAP and Planck). In this way, we can test our cosmological models using the CMB quadrupole.

Note that the proposed test directly compares the CMB quadrupole transfer functions that depend on the universe's expansion history after the initial density field, which is the origin of the cosmic variance, has been estimated and fixed. Therefore, we do not suffer from the cosmic variance uncertainty that is large for the CMB quadrupole measurement compared with the instrumental noise. Using simple simulations, this study aims to show that one can test for dark energy models in this way beyond the cosmic variance.

\section{Methodology}
We start with the Stokes parameters $Q(\vec{x})$ and $U(\vec{x})$ of the CMB photons induced by the primordial CMB quadrupole at a cluster position $\vec{x}$ \cite{2018JCAP...04..034D}: 
\begin{equation}
 Q(\vec{x})\pm iU(\vec{x})=-\frac{\sqrt{6}}{10}\tau\sum_m {_{\pm 2}}Y_{2
  m}(\hat{x})a_{\rm 2m}(\vec{x})~.
\label{eq:QiU}
\end{equation}
Here, $\tau$ is the optical depth of the cluster, and $a_{\ell m}(\vec{x})$ are the coefficients of the spherical harmonic expansion of the CMB temperature field at position $\vec{x}$,
\begin{equation}
 \frac{\delta T}{T}(\vec{x},\hat{n},\eta)=\sum_{\ell m} a_{\ell
  m}(\vec{x})Y_{\ell m}(\hat{n})~,
\end{equation}
with $\eta=\eta_0-|\vec{x}|$ being the conformal time of the scattering events.  The coefficients of the spherical harmonic expansion are expressed using the Fourier modes as
\begin{equation}
 a_{\ell m}(\vec{x}) = (-i)^\ell (4\pi) \int d^3 k
  e^{i\vec{k}\cdot\vec{x}} \Delta_\ell(\vec{k},\eta) Y^\ast_{\ell m}(\hat{k})~,
\end{equation}
where $\Delta_\ell(\vec{k},\eta)$ is the Legendre expansion coefficients of the CMB photon distribution. We may further expand $\Delta_\ell(\vec{k},\eta)$ as
\begin{equation}
 \Delta_\ell(\vec{k},\eta) = \Delta_\ell(k,\eta) {R}_{\rm ini}(\vec{k})~,
\label{eq:transfer}
\end{equation}
where $\Delta_\ell(k,\eta)$ are the linear transfer functions that depend on a cosmological model, and ${R}_{\rm ini}(\vec{k})$ is the initial curvature perturbation with the ${P}(k)$ variance.

\begin{figure}
    \centering
    \includegraphics[width=1.\hsize]{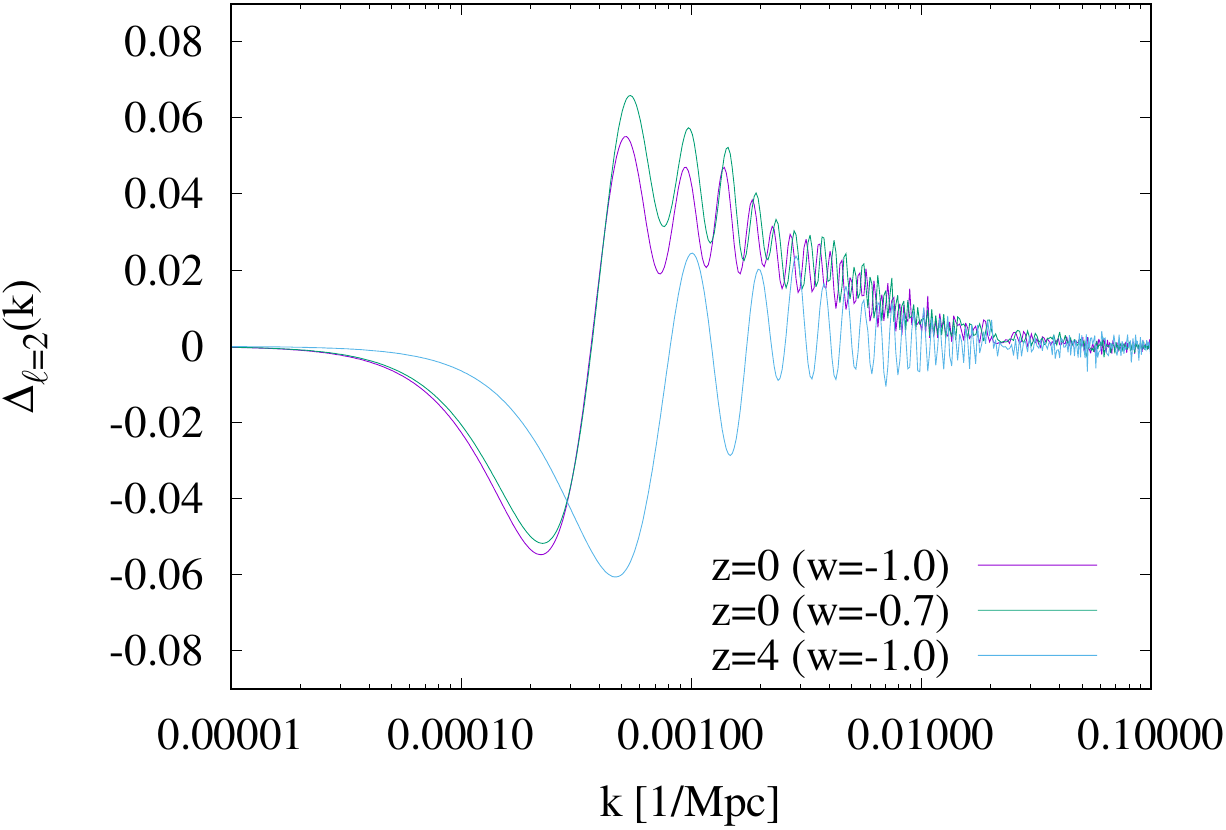}
    \caption{Transfer function of the CMB quadrupole, $\Delta_{\ell=2}(z,k)$ for $(z,w)=(0,-1)$, $(0,-0.7)$ and $(4,-1)$ as indicated in the figure.}
    \label{fig:l2transfer}
\end{figure}

In our simulation, we generated the transfer functions in Eq.~(\ref{eq:transfer}) using CAMB a publicly available code \cite{Lewis:1999bs}. Figure (\ref{fig:l2transfer}) shows the transfer functions at $z=0$ and $z=4$ with different equation-of-state parameters of dark energy, $w\equiv P_{\rm DE}/\rho_{\rm DE}$. Comparing the transfer functions at $z=0$ and $z=4$, the ISW contribution was apparent at wavenumbers $k\approx 10^{-3}$ Mpc$^{-1}$. As also shown in the figure, the larger $w$ parameter ($w=-0.7$) induced the larger ISW effect at $z=0$. 

We generated the initial curvature perturbation ${R}_{\rm ini}(\vec{k})$ as a random Gaussian variable with a dimensionless power spectrum:
\begin{equation}
 {\cal P}(k)=\frac{k^3}{2\pi^2} P(k) = A_s \left(\frac{k}{k_\ast}\right)^{n_s-1}~,
  \label{eq:equation_Pk}
\end{equation}
where $k_\ast=0.05$ Mpc$^{-1}$, $A_s=2.1\times 10^{-9}$ and $n_s = 0.96$. 
For the other cosmological parameters, we fixed them to the standard $\Lambda$CDM values, i.e., $\Omega_bh^2 = 0.0226$, $\Omega_ch^2=0.112$, $\Omega_\nu h^2 = 0.00064$, $h=0.7$ where $\Omega_bh^2$, $\Omega_ch^2$ and $\Omega_\nu h^2$ are the baryon density, cold dark matter density, and neutrino density, respectively, and $h$ is the normalized Hubble parameter.

\begin{figure}
    \centering\includegraphics[width=1.\hsize]{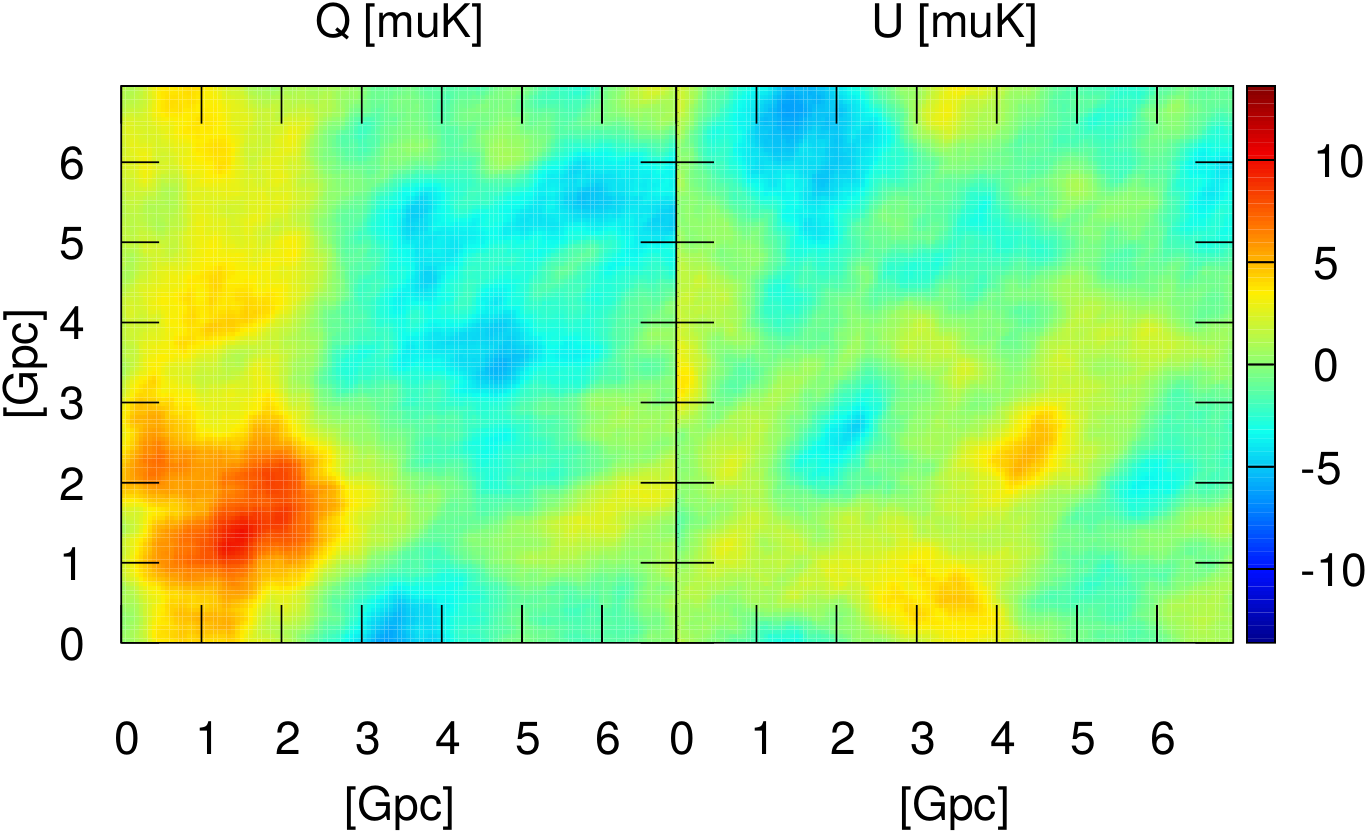}
    \caption{Example realization for the Stokes Q (left) and U (right) maps from the CMB quadrupole defined in Eq.~(1) at $z=0$. The signals are correlated on very large scales \cite{2004PhRvD..70f3504P}. We assume $\tau=1$ in Eq.~(\ref{eq:QiU}) for simplicity.}
    \label{fig:slice_QU}
\end{figure}

The procedure of our methodology is as follows:
\begin{itemize}
\item[1.] We generated ${R_{\rm ini}}(k_i)$ according to the power spectrum given by Eq.~(\ref{eq:equation_Pk}). Following our previous paper \cite{2016MNRAS.460L.104L}, we sampled the Fourier mode in angular directions on the Healpix grid with $N_{\rm side}=8$, and 60 modes uniformly in a logarithmic space in a radial direction from $k=10^{-6}$ to $10^{-1}$ Mpc$^{-1}$. Thus, the total number of the Fourier mode is thus $n_k = 46,080$.
    
\item[2.] Using the generated $R_{\rm ini}(k_i)$, we simulated $Q_{\rm
	  fid}(x_i)$ and $U_{\rm fid}(x_i)$ at the cluster positions on
	  our past lightcone assuming a true cosmological model (in our
	  case, $w=-1$, the $\Lambda$CDM model). We then added Gaussian
	  noises to $Q_{\rm
	  fid}(x_i)$ and $U_{\rm fid}(x_i)$ with a variance
	  $\sigma_{\rm pol}$. Figure~(\ref{fig:slice_QU}) shows a realization of the $Q$ and $U$ maps. We considered $6000$ randomly distributed clusters $N_{\rm cluster} = 6000$ from $z=0$ to $z=2$. We also calculated the CMB quadrupole at the origin $a_{2m}^{\rm true}(0)$.

\item[3.] As a backword problem, we estimated the initial curvature perturbation, ${R}_{\rm ini}(k_i)$,  by fitting to the $Q$ and $U$ map minimizing \cite{2016MNRAS.460L.104L}
    \begin{eqnarray}
    f &=& \sum_i^{N_{\rm cluster}} \left[\frac{(Q(x_i)-Q_{\rm fid}(x_i))^2}{2\sigma_{\rm pol}^2} 
    + \frac{(U(x_i)-U_{\rm fid}(x_i))^2}{2\sigma_{\rm pol}^2} \right]
    \nonumber \\                                           \
    &+&\sum_{j}^{n_k} \frac{{R}^2_{\rm ini}(k_j)}{2P(k_j)}~, \nonumber
    \end{eqnarray}
where $\sigma_{\rm pol}$ describes the uncertainty in observing $Q$ and $U$ at the cluster position. 
Here the first two terms are for the chisquare-minimization for $Q(x_i)$
	  and $U(x_i)$, which depend on the initial curvature
	  perturbation $R_{\rm ini}(k_i)$, using a fiducial mock data
	  $Q_{\rm fid}(x_i)$ and $U_{\rm fid}(x_i)$. In a real
	  application, $Q_{\rm fid}(x_i)$ and $U_{\rm fid}(x_i)$ will be the observables and we
	  search or tune $R_{\rm ini}(k_i)$ to find $Q(x_i)$ and $U(x_i)$ that minimize the function $f$.
	  The third term represents the
	  Gaussian prior on the initial curvature perturbation $R_{\rm ini}(k_i)$ with a variance $P(k_i)$. 
 We repeated this process for various equation of state parameters $w$.
    
\item[4.] After obtaining the estimated initial curvature perturbation $R^{\rm est}_{\rm ini}(k_i)$, we calculated the CMB quadrupole at the origin, $a^{\rm est}_{2m}(0)$, using $R^{\rm est}_{\rm ini}(k_i)$ and compared it to $a^{\rm true}_{2m}(0)$ calculated in the second step.
    
\item[5.] Go back to Step $1$ with different initial perturbations and cluster positions and repeat the procedure a hundred times.
\end{itemize}
The true and estimated CMB quadrupoles,  $a^{\rm true}_{2m}(0)$ and
$a^{\rm est}_{2m}(0)$, should coincide within the methodological
statistical uncertainty if we use the correct transfer function in
Step 3 above. In other words, we can constrain cosmological models if
$a^{\rm true}_{2m}(0)$ and $a^{\rm est}_{2m}(0)$ differ beyond the
methodological statistical uncertainty.

\section{Result}
\subsection{Instrumental error in the CMB quadrupole measurement and
    methodological statistical uncertainity}
The reconstruction of the local CMB quadrupole using the remote quadrupole information is not perfect due to the limited number of galaxy clusters available and the polarization measurement errors. We only show herein the results for the case where $6000$ galaxy clusters were used with $\sigma_{\rm pol}/\tau=10^{-2}$, $3\times10^{-2}$ and $10^{-1}$ $\mu$K. Although we argued in our previous paper that $300$ clusters would be sufficient to estimate the local quadrupole, we found that the number is not enough for the investigation of dark energy. A detailed study will be presented in our future work \cite{Sumiya.et.al.}.

We first defined the statistical uncertainity in our methodology
$\sigma_{\rm method}$ by an evaluation similar with $C_\ell$:
\begin{equation}
    \sigma^2_{\rm method} = \frac{1}{N}\sum_{i=1}^{N}\frac{1}{5}
    \left[
    |\Delta a_{20}|^2+2|\Delta a_{21}|^2+2|\Delta a_{22}|^2
    \right]~,
\end{equation}
where, $\Delta a_{2m} = a_{2m}^{\rm est}(w=-1) - a_{2m}^{\rm true}(w=-1)$, and $N$ is the number of simulations.

The variance $\sigma_{\rm method}$ may depend on the simulation setup.
It directly depends on $\sigma_{\rm pol}$, and at the same time, it
also depends on the detailed distribution of the cluster, such as their
positions, number density, redshift range, and so on. Thus, it is
difficult to find the dependency analytically. Therefore, we numerically
evaluate $\sigma_{\rm method}$ for some $\sigma_{\rm pol}$ values.
In our fiducial setup with $N_{\rm cluster}=6000$, $\sigma_{\rm pol}/\tau=10^{-2}$ $\mu$K, $N_{\rm side}=8$, and $n_{\rm kmode}=60$, we found
\begin{equation}
\sigma_{\rm method} \simeq 2.4 \times 10^{-8}~,
\end{equation}
and we had $\sigma_{\rm method}\simeq 5.9 \times 10^{-8}$ for
$\sigma_{\rm pol}/\tau=3\times 10^{-2}$ $\mu$K.

The next generation CMB satellite, namely LiteBIRD, will reach $\sim 2.0 ~\mu$Karcmin sensitivity \cite{2020SPIE11443E..2FH}.  The angular scale of the CMB quadrupole is $\sim 90^\circ = 5400$ arcmin. We may expect that, in an ideal case, the CMB quadrupole viewed from us can be measured at the following level:
\begin{equation}
    \sigma^{\rm LiteBIRD}_{\ell=2} \simeq \frac{2.0}{5400}
    \simeq 3.7\times 10^{-4} \mu{\rm K}~.
\end{equation}
In the dimensionless units normalized by the mean CMB temperature, we had $\sigma_{\ell=2} \simeq \sigma^{\rm LiteBIRD}_{\ell=2}/(2.725 \times 10^6 \mu{\rm K})=1.4\times 10^{-10}$. Variance $\sigma_{\ell=2}$ is two orders of magnitude smaller than the error associated with our methodology; thus, we may neglect it.
The conclusion here is that the uncertainty in measurements of local quadrupoles will not limit the precision of this method, but the methodological, statistical uncertainty associated with measurements of remote quadrupoles.

\subsection{Constraints on the dark energy parameter $w$}
To make a simple statistical inference, we define the chi-squared statistics as follows:
\begin{eqnarray}
\chi^2(w)
 &=&\frac{1}{\sigma^2_{\rm method}}\left(
		       |\Delta a_{20}|^2+2|\Delta a_{21}|^2+2|\Delta a_{22}|^2
		      \right)~,
\label{eq:chisq}
\end{eqnarray}
where, $\Delta a_{2m}=a^{\rm est}_{2m}(w)-a^{\rm true}_{2m}(w=-1)$.
We had real and imaginary parts in each $\Delta a_{2m}$. Accordingly, we assigned the variance $\sigma^2/2$ for $m=1$ and $m=2$ components and $\sigma^2$ for the $m=0$ component because it had only the real part. 

The chi-square defined in Eq.~(\ref{eq:chisq}) is a measure of the
difference in the goodness-of-fit between different models. For example,
if we use the correct dark energy parameter $w$ that coincides with the
input value, i.e., $\Delta a_{2m}=a^{\rm est}_{2m}(w=-1)-a^{\rm
true}_{2m}(w=-1)$, $\chi^2(w=-1)$ should follow the chi-square
distribution with a degree of freedom of five. If we use incorrect
dark energy parameters ($w \neq -1$), we have larger $\chi^2$ values, and
thereby we may 
constrain the dark energy parameter. Following the standard $\chi^2$ method,
we expect $1\sigma$ and $2\sigma$ constraints for $\Delta \chi^2=1$ and
$4$, respectively.

\begin{figure}
    \centering
    \includegraphics[width=\linewidth]{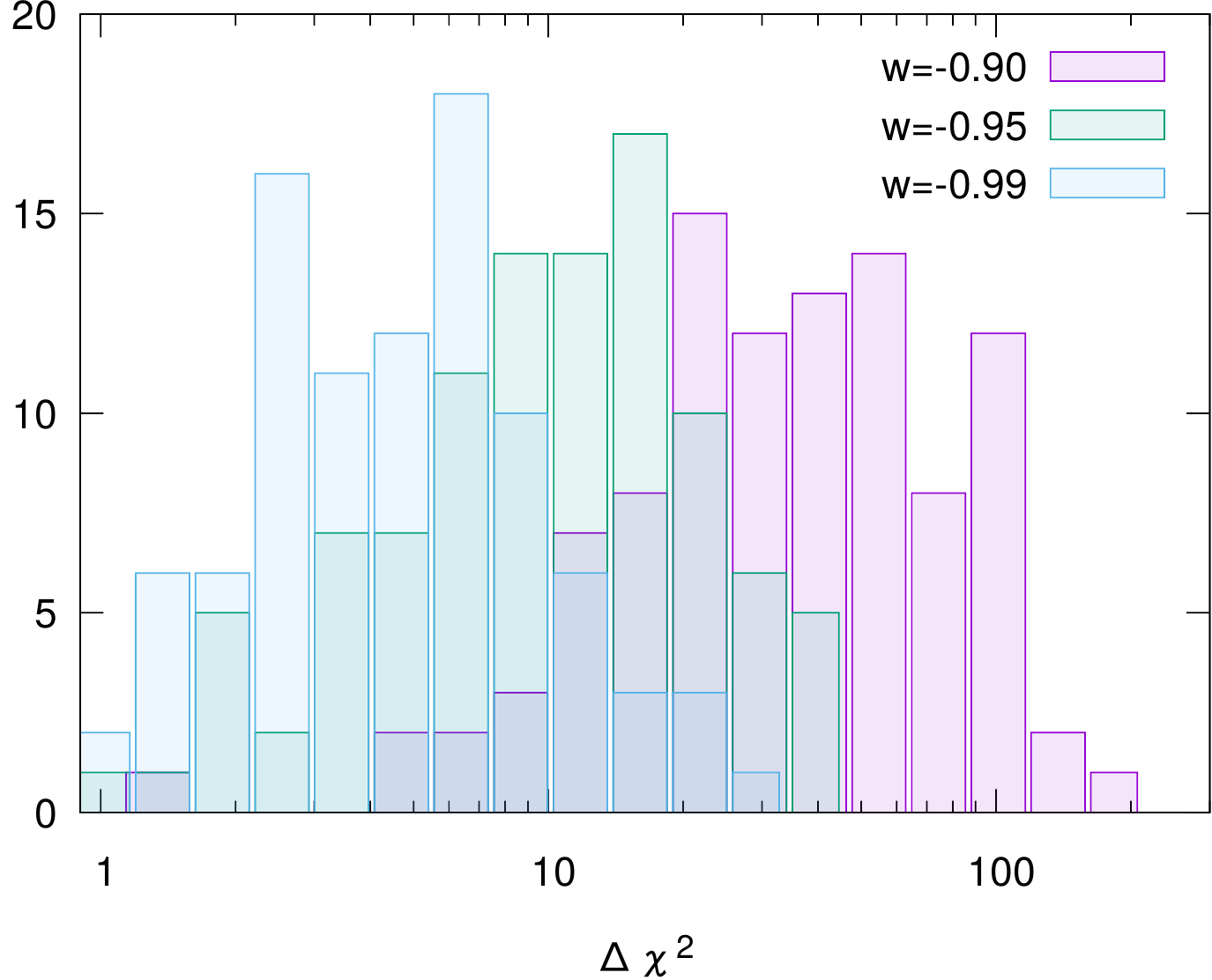}
 \caption{Distribution of $\Delta \chi^2=\chi^2(w)-\chi^2(w=-1)$ where $\chi^2$ is defined in Eq.~(\ref{eq:chisq}). We assumed that $\sigma_{\rm pol}/\tau=10^{-2}$ $\mu$K. The simulation setup was $N_{\rm cluster} = 6000$, $N_{\rm side}=8$, and $n^{\rm radial}_{\rm kmode}=60$.}  \label{fig:chisq_dist_sig1e-2}
\end{figure}

Compared with the $w=-1$ model in the hundred simulations, we found that $\langle\Delta \chi^2\rangle\approx 1.56$, $9.86$, and $47.0$ for the
$w=-0.99$, $-0.95$ and $w=-0.90$ models, respectively. Figure~(\ref{fig:chisq_dist_sig1e-2}) illustrates the $\Delta\chi^2$ distribution in our simulation.
The statistical test significance strongly depends on $\sigma_{\rm pol}$.
If we assume less sensitivity in the polarization measurement as $\sigma_{\rm pol}/\tau=3\times 10^{-2}$
$\mu$K, the delta chi--squared
reduces to $\langle\Delta \chi^2\rangle\approx 0.74$, $4.0$ and $13.1$
for $w=-0.99$, $0.95$ and $-0.90$, respectively.
We will present a more detailed analysis in our future work~\cite{Sumiya.et.al.}.

\section{summary and discussion}

In this study, we proposed a new method for use in observational cosmology using CMB observations. The conventional way in observational cosmology has been to obtain summary statistics, such as the variance of cosmological density fluctuations, from observations at various times in the history of the universe and compare them with theory. However, this method cannot escape the cosmic variance arising from the fact that there is only one observable universe. Instead of using summary statistics, we estimated and fixed the initial density fluctuations realized in our universe using the polarization of distant galaxy clusters in the past and determined the time evolution of the density fluctuations by considering how the density fluctuations look today using the local quadrupole of the CMB. This process corresponds, in effect, to multiple density fluctuation observations in the universe, hence the title of the paper.

As a working example, we considered $w$CDM cosmology and investigated
how sensitive this method is to the equation of the state parameter $w$,
assuming that the $\Lambda$CDM cosmology is the correct cosmological
model.
In Sec. III, we found that the expected $\Delta \chi^2$ value compared to the $w=-1$ model is as large as $9.86$ for $w=-0.95$. Therefore, we concluded that
one may be able to achieve $5$\% precision for constraint on $w$ if the CMB polarization, which is caused by the remote quadrupole at $6000$ galaxy cluster positions up to redshift $z=2$, is accurately and precisely measured below the noise level of $\sigma_{\rm pol}\lesssim 10^{-2}\tau$ $\mu$K, where $\tau$ is the optical depth of the cluster.

The expected signal of the typical cluster polarization contributions caused by the quadrupole ranges from $0.02$ to $0.1$ $\mu$K \cite{2003PhRvD..67f3505C}.  Therefore, detecting signals from individual galaxy clusters would require a detector with this level of sensitivity at arc-minute angular scales. In reality, however, this is not necessary \cite{2014PhRvD..90f3518H}. As we have already discussed, the polarization signals from galaxy clusters are correlated on large scales; hence, rather than detecting the signals of individual galaxy clusters, we only need to catch the signals aligned on large scales \cite{2017PhRvD..96l3509L} (Fig.~(\ref{fig:slice_QU})). As shown herein, $6000$ locations between redshifts $z=0$ and $z=2$ are sufficient. As a rough estimate, the volume of the universe up to redshift $z=2$ is approximately $600$ Gpc$^3$, which means that we can average the signals from the clusters of the galaxies over a volume of $0.1$ Gpc$^3$. 
At $z\lesssim 1$ we may expect approximately $10^3$ clusters in that volume with $\sim 10^{14}M_\odot$ halo mass.

The simulations we performed are idealized in many aspects. In our simulation, the magnitude of the galaxy cluster polarization was assumed to be known, and the noise in the polarization observations was supposed to be very small. Except for the CMB quadrupole, we also ignored the polarization sources, such as those caused by the kSZ effect and the primordial polarization of the background CMB \cite{1999MNRAS.310..765S,2012ApJ...757...44R}, although separation would be possible based on their different frequency spectra \cite{2003PhRvD..67f3505C,1997ApJ...482..577D}. In addition, the instrumental noise-limited quadrupole measurement of the CMB and the signal separation from other contaminants would be an experimental challenge due to the correlated noise in the time-ordered CMB data and the galactic foregrounds.

Although our work is preliminary, we have shown that future cluster polarization measurements combined with the local CMB quadrupole measurement would offer a powerful probe for the nature of dark energy. More importantly, this method is qualitatively distinct from conventional methods based on summary statistics, which always suffer from cosmic variance errors. A more detailed study in a more realistic setting is necessary.

 \begin{acknowledgments}
  One of the authors (KI) would like to thank D. Huterer and A. Cooray
  for helpful communication.
This work is supported in part by the JSPS grant numbers 18K03616,
17H01110 and JST AIP Acceleration Research Grant JP20317829 and JST
FOREST Program JPMJFR20352935 (K.I.), and the Ministry of Science and Technology (MOST) of Taiwan, Republic of China, under Grants No. MOST 109-2112-M-032-006 (G.C.L.)
 \end{acknowledgments}



\bibliographystyle{apsrev4-2}
\bibliography{draft}

\end{document}